\documentclass[useAMS,usenatbib]{mnras}
\pdfoutput=1
\usepackage{epsfig}
\usepackage{graphicx}
\usepackage{amssymb} 
\usepackage{amsmath}

\title[The Silhouette Jet HH~1019]{The Dusty Silhouette Jet HH~1019 in
  the Carina Nebula}

\author[Reiter et al.]{Megan Reiter$^1$\thanks{Email:
    mreiter@umich.edu}, Megan M. Kiminki$^2$, Nathan Smith$^2$, 
  and John Bally$^3$ \\ 
  $^1$University of Michigan, 1085 S. University Ave., Ann Arbor, MI 48109, USA \\
  $^2$Steward Observatory, 933 N. Cherry Ave., Tucson, AZ 85721, USA \\
  $^3$Center for Astrophysics and Space Astronomy, University of Colorado, 389 UCB, Boulder, CO 80309, USA }

\begin{document}

\date{Accepted 2017 February 10. Received 2017 February 10; in original form 2016 November 4}
\pagerange{\pageref{firstpage}--\pageref{lastpage}} \pubyear{2016}
\def\arcdeg{\degr}
\maketitle
\label{firstpage}

\begin{abstract}

  We report the discovery in {\it Hubble Space Telescope} ({\it HST})
  images of the new Herbig-Haro jet, HH~1019, located near the Tr~14
  cluster in the Carina Nebula.  Like other HH jets in the
  region, this bipolar collimated flow emerges from the head of a dark
  dust pillar.  However, HH~1019 is unique because -- unlike all
  other HH jets known to date -- it is identified by a linear chain 
  of dark, dusty knots that are seen primarily in silhouette against
  the background screen of the H~{\sc ii} region.
Proper motions confirm that these dark condensations move along the jet axis at high speed. 
  [S~{\sc ii}] emission traces a highly collimated jet that is spatially coincident with these dust knots. 
  The high extinction in the body of the jet suggests that this outflow has lifted a large amount of dust directly from the disk, although it is possible that it has entrained dust from its surrounding protostellar envelope before exiting the dust pillar.
If dust in HH~1019 originates from the circumstellar disk, this provides further evidence for a jet launched from a range of radii in the disk, including those outside the dust sublimation radius. 
  HH~1019 may be the prototype for a new subclass of dusty HH objects 
  seen primarily in extinction against the background
  screen of a bright H~{\sc ii} region.
  Such jets may be common, but difficult to observe because they require the special condition of a very bright
  background in order to be seen in silhouette.
\end{abstract}


%%%%%%%%%%%%%%%%%%%%%%%%%%%%%%%%%%%%%%%%%%%%%%%%%%%%%%%%%%%%%%%%%%%%%%%%%%
\section{INTRODUCTION}

Protostellar jets accompany active disk accretion onto forming stars.
While the detailed physics of jet launch and collimation remain uncertain, accretion energy must power the outflow.
Different theories place the jet-launch footpoint at different radii in the disk, but all must explain the observed onion-like velocity structure and high degree of collimation seen at large distances from the protostar \citep[see, e.g.][]{fer06}.

\begin{figure*}
\centering
  \includegraphics[trim=17.5mm 15mm 15mm 15mm,angle=0,scale=0.6175]{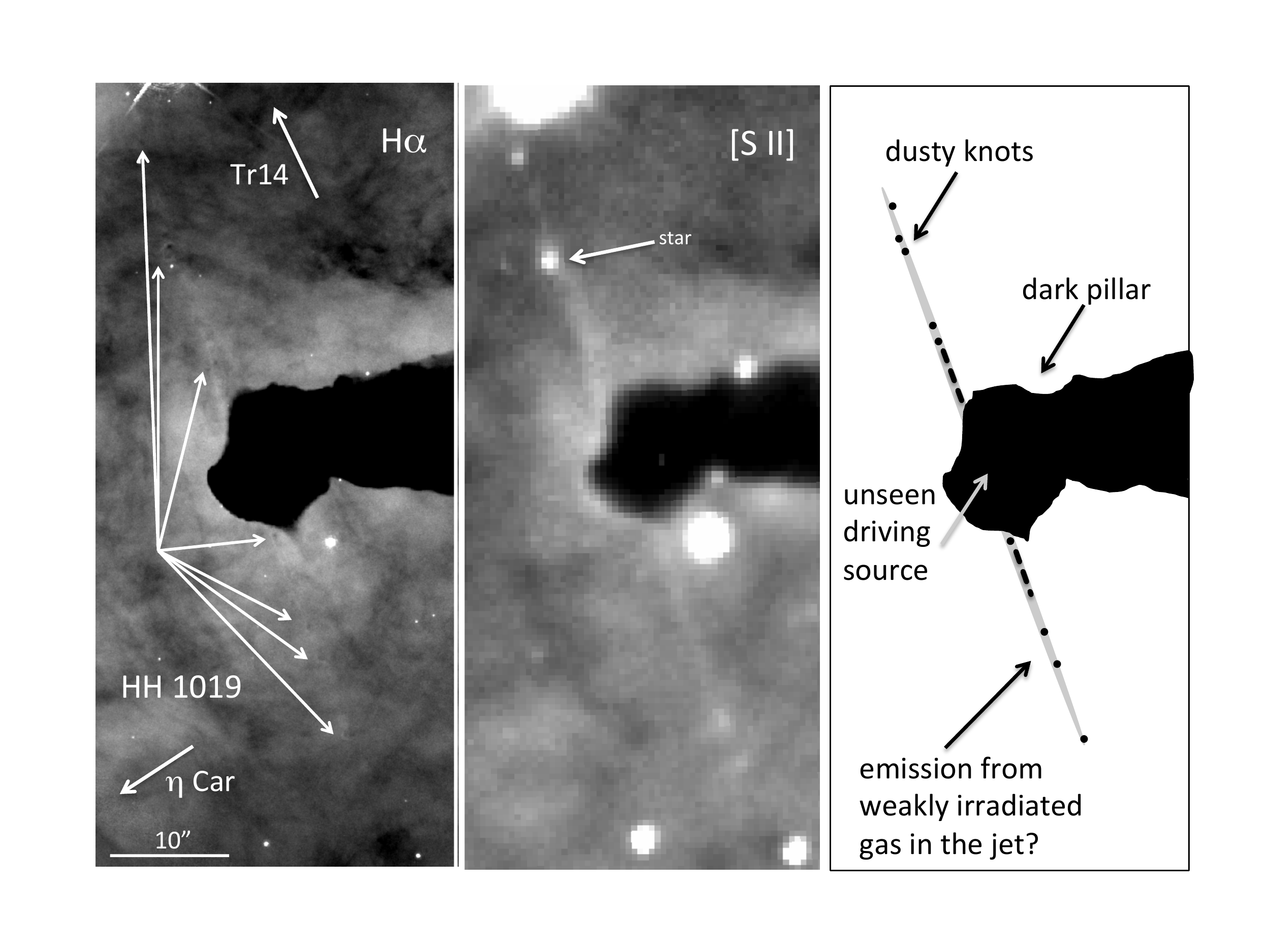} 
\caption{
  \textit{Left:} \emph{HST}/ACS H$\alpha$ image of HH~1019 that shows the chain of dusty knots that make up the jet seen in silhouette against the bright background of the H~{\sc ii} region.  
  \textit{Middle:} [S~{\sc ii}] emission clearly traces a narrow, collimated jet even though the detailed structure is unresolved in the seeing-limited ground-based image.
  \textit{Right:} Cartoon showing the structure of HH~1019. Dark, dusty knots (black) trace out a collimated bipolar jet emerging from an opaque dust pillar. Faint H$\alpha$ and [S~{\sc ii}] emission (gray) trace the weakly irradiated gas in the jet. 
 }\label{fig:hh1019_intro} 
\end{figure*}

Jet composition may offer an indirect probe of the jet launch region.
Provided that shocks and environmental interaction have not significantly altered the jet, its composition will reflect the material at the launch radius in the disk.
In this picture, an outflow launched from a large range of disk radii, as in a disk wind, may include matter both inside and outside the dust sublimation radius \citep[see, e.g.][]{aa11}.
In contrast, an X-wind that is launched near the corotation radius will sample the more highly excited material near the star. 
For more massive sources, higher luminosities may evaporate grains near the star so that material launched from the inner disk may be dust-free \citep[e.g.][]{kam09}.

Dust in protostellar jets, especially in the body of the jet between shock fronts, is taken as evidence that at least a portion of the jet was launched outside the dust sublimation radius \citep[see, e.g., discussion in][]{aa11}. 
However, jets may entrain outflows as they pass through the natal cloud.
Indeed, dust has been observed in molecular outflows \citep[e.g.][]{shi00,chi01,gue03}, making it unclear whether the dust originates from the disk or if it is merely entrained from the environment. 
Shocks in the jet may also destroy dust.
However, \citet{mou00} suggest that shock destruction is inefficient, allowing a significant fraction of dust grains to survive.

Despite the importance of dust in outflows, few observations provide \textit{direct} evidence for dust.
\citet{chi01} detect millimeter continuum emission from a few Herbig-Haro (HH) objects, the shock-excited nebulosities associated with protostellar jets that are typically identified via their optical and IR emission \citep{her50,her51,har52,har53}.
These sources are associated with CO outflows and young driving sources, leading \citet{chi01} to suggest that dust is heated in the interaction between the jet and the environment. 
Extended (sub)-mm continuum emission that traces the bipolar outflow lobes has been detected in the L~1157 molecular outflow \citep{shi00,gue03}. 
Many molecular lines contribute to the flux observed at 1.3~mm and 850~\micron, however \citet{gue03} conclude that line emission can only account for $\lesssim 50$\% of the continuum flux. 
In addition, the spectral index of the 1.3~mm and 850~\micron\ emission is consistent with thermal dust emission. 
However, these observations have an angular resolution of several arcsec \citep[in the case of L~1157, 12\arcsec\ corresponding to 5280~AU a distance of 440~pc,][]{gue03}, making it unclear whether the dust in the outflow is entrained from the environment.

\begin{figure*}
\centering
$\begin{array}{c}
  \includegraphics[angle=0,scale=0.575]{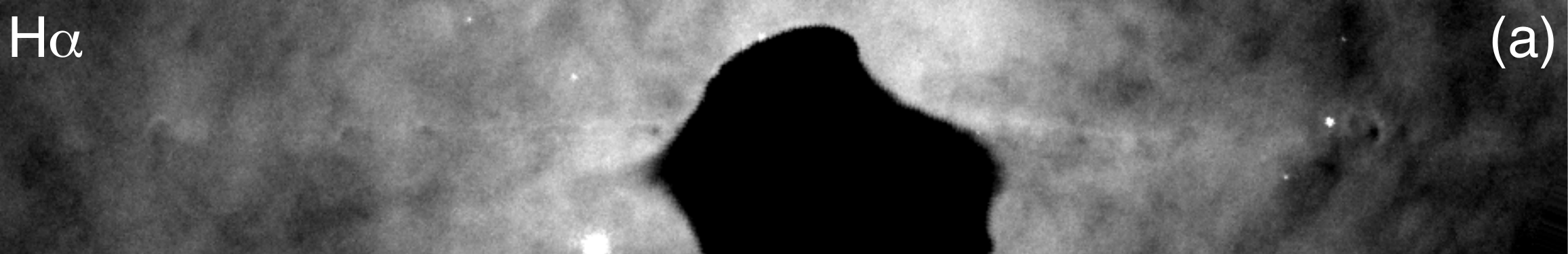} \\
  \includegraphics[angle=0,scale=0.575]{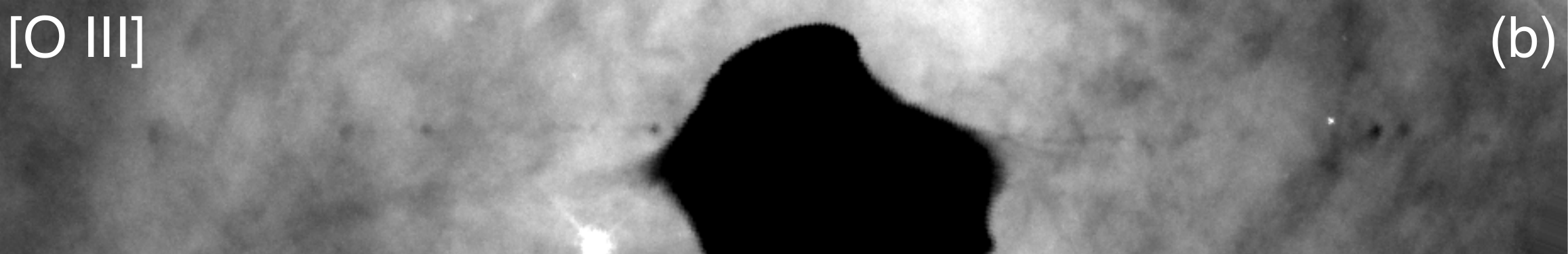} \\
  \includegraphics[angle=0,scale=0.575]{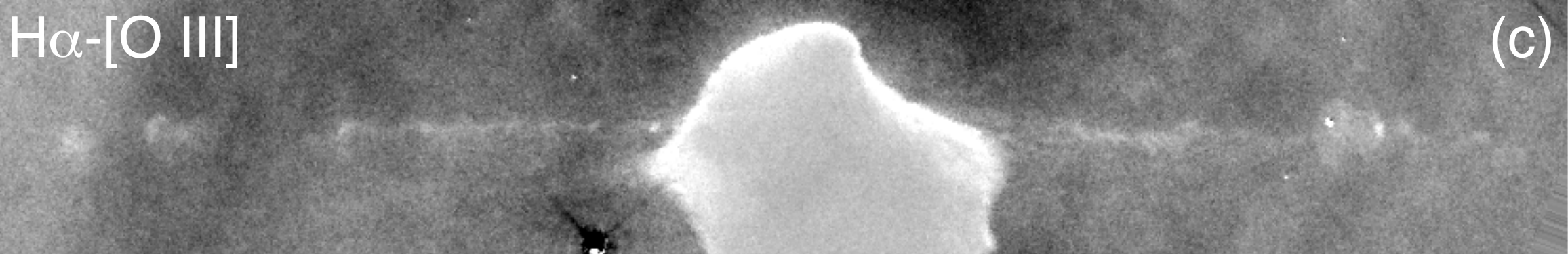} \\
  \includegraphics[angle=0,scale=0.575]{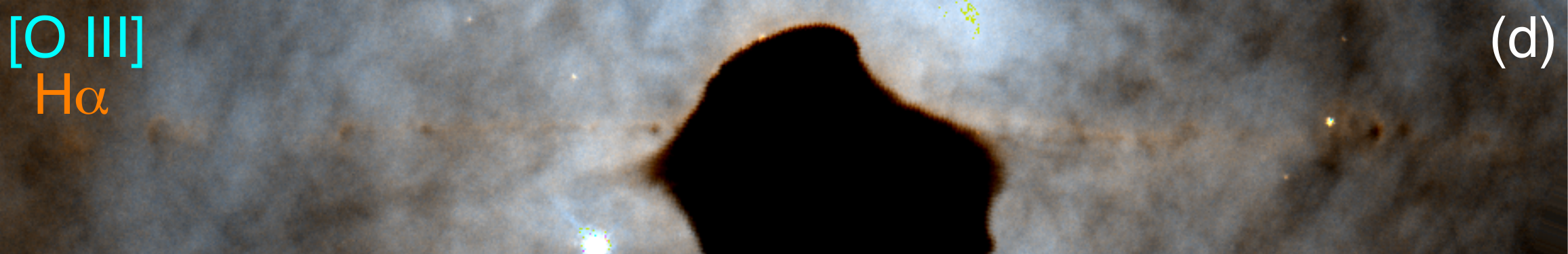} \\
\end{array}$
\caption{
  (a) Zoomed in view of HH~1019 in the \emph{HST}/ACS H$\alpha$ image. The jet emerges just above the widest part of the pillar and extends through the middle of the image. 
  (b) Jet knots identified in the H$\alpha$ image are also seen in silhouette in a \emph{HST}/ACS [O~{\sc iii}] image. Streams of silhouetted material can also be seen emerging on either side of the pillar head. 
  (c) An H$\alpha$-[O~{\sc iii}] image suppresses nebular emission, clearly showing a collimated, bipolar jet. 
  (d) Color image made combining H$\alpha$ and [O~{\sc iii}] from \emph{HST}/ACS. Dusty jet knots, like the pillar, are dark against the bright nebulosity of the H~{\sc ii} region. 
 }\label{fig:hh1019_hst} 
\end{figure*}

Most studies infer the presence of dust in the highly collimated jet itself from the gas-phase depletion of refractory elements like iron or calcium relative to solar abundance \citep[e.g.][]{aa11,nis05,pod11}. 
One exception is \citet{smi05} who report thermal dust emission at 11.7~\micron\ in a few jets in Orion. 
With higher angular resolution ($<1$\arcsec) than the millimeter observations, \citet{smi05} show that the mid-IR emission is spatially associated with the ionized gas in the \textit{body} of at least one jet (HH~513) that was likely not entrained from the surrounding molecular cloud. 
Such clear evidence for dust in the jet body argues strongly for dust being launched into the flow rather than created in post-shock cooling zones.  
However, few comparable examples of dust in the jet body exist.

In this paper, we present the discovery of another member of this rarefied group of jets with direct evidence for dust. 
Unlike other Herbig-Haro (HH) jets, HH 1019 in the Carina Nebula appears to be composed of a chain of dark, dusty knots seen \textit{in silhouette} against the background screen of the H~{\sc ii} region (see Figure~\ref{fig:hh1019_intro}).  
UV photons from nearby Tr14 and Tr16 illuminate the edges of dusty knots.
This is distinct from the illumination of the body observed in most of the other 40 HH jets discovered in the same H$\alpha$ imaging survey of Carina \citep{smi10}.
The high mass-loss rate inferred for those jets implies intermediate-mass driving sources, suggesting that the HH jets in Carina offer a unique view of the underlying \textit{jets} that may be common in the formation of higher-mass stars \citep{smi10,ohl12,rei13,rei16}.
The key difference is that HH~1019 is seen in silhouette, revealing a dusty jet that would remain unseen without the bright background screen. 
As such, HH~1019 is an important target for understanding the origin and evolution of dust in protostellar jets.

%%%%%%%%%%%%%%%%%%%%%%%%%%%%%%%%%%%%%%%%%%%%%%%%%%%%%%%%%%%%%%%%%%%%%%%%%%%%%%
\section{OBSERVATIONS}

HH~1019 was first imaged as part of a larger H$\alpha$ imaging survey with the Advanced Camera for Surveys (ACS) on the \emph{Hubble Space Telescope} (\emph{HST}) presented and described in detail by \citet{smi10}.
However, HH~1019 was not reported as one of the 40 HH jets discovered in the survey. 
In this paper, we include part of the Tr14 mosaic from \citet{smi10} that shows HH~1019 and its natal dust pillar. 
The position of HH~1019 is RA=10:44:00.3 and DEC=-59:36:15. 
The first-epoch image was obtained 2005 Jul 17 \citep[as part of programmes GO-10241 and GO-10475; see][]{smi10}. 

We also present a second epoch \emph{HST}/ACS image of the same region obtained 2015 Jun 28 (programme GO-13390, PI: N.\ Smith).
Both epochs were observed in the F658N filter (which transmits both H$\alpha$ and [N~{\sc ii}] $\lambda 6583$) with an exposure time of $1000$~s.
Comparing the two epochs allows us to measure proper motions over a time baseline of $\sim 10$~yr.
Ideally, the second epoch observation will duplicate the setup of the first epoch as closely as possible.
However, a change in the guide stars required a 180$^{\circ}$ rotation between epochs. 
This rotation introduces an additional systematic uncertainty in the proper motion measurement of HH~1019.

To align and stack the images, we follow the procedure described in detail in \citet{rei15a,rei15b}, which is an adaptation of the method of \citet{anderson2008a,anderson2008b}, \citet{andersonvandermarel2010}, and \citet{sohn2012}. 
Images of each tile are aligned to a common reference frame using stellar positions measured with point spread function (PSF) photometry.
The final images have 50-mas pixels and are aligned so that north is up. 
The alignment reference frame is the average position of the stars in the image, and thus is not tied to an external zero point.
Jet proper motions are therefore in the frame of the Carina Nebula.
Where the alignment of the stars and gas disagree, we assume that large nebular structures like dust pillars are stationary. 
We apply a linear shift to the second-epoch stacked images so that the relative motion of the pillar is zero. 
To ensure that we are measuring real motion despite these systematics, we only report features faster than $\sim 25$~km~s$^{-1}$.

To measure the proper motions of knots in the jet, we use the modified cross-correlation technique developed by \citet{cur96,har01,mor01}.
As described in detail in, e.g., \citet{rei14,kim16}, we identify knots that do not change significantly in brightness and morphology between epochs. 
We subtract a median-filtered image (with a 25~pix kernel) to suppress background emission, then extract the jet knot in a small box optimized for each feature. 
We apply a series of small shifts of this box between the two epochs.
For each shift, we compute the total of the square of the difference between the two images. 
The minimum value of the resulting array gives the offset required to align a feature in the two images.

In addition, we obtained a narrowband [O~{\sc iii}] image (F502N) taken with \emph{HST}/ACS from the \emph{HST} archive\footnote{http://archive.stsci.edu/prepds/carina/wfc3/}. 
This mosaic was obtained as a parallel field as part of PID 12050 (PI M. Livio) during 1-2 February 2010. 
Like the complementary \emph{HST}/WFC3 data, the drizzled [O~{\sc iii}] image is a $\sim 6\arcmin \times 6\arcmin$ mosaic of four overlapping pointings. 
The total exposure time is 30600~s.

We also include a narrowband [S~{\sc ii}] $\lambda\lambda 6717, 6731$ image from MOSAIC2 imager on the Blanco 4~m at CTIO.
The [S~{\sc ii}] image was obtained on 18 December 2001 and has a total integration time of 480~s. 
Additional details about image and data reduction are presented in \citet{smi03,smi05b}.

%%%%%%%%%%%%%%%%%%%%%%%%%%%%%%%%%%%%%%%%%%%%%%%%%%%%%%%%%%%%%%%%%%%%%%

\begin{figure}
\centering
  \includegraphics[angle=0,scale=0.5]{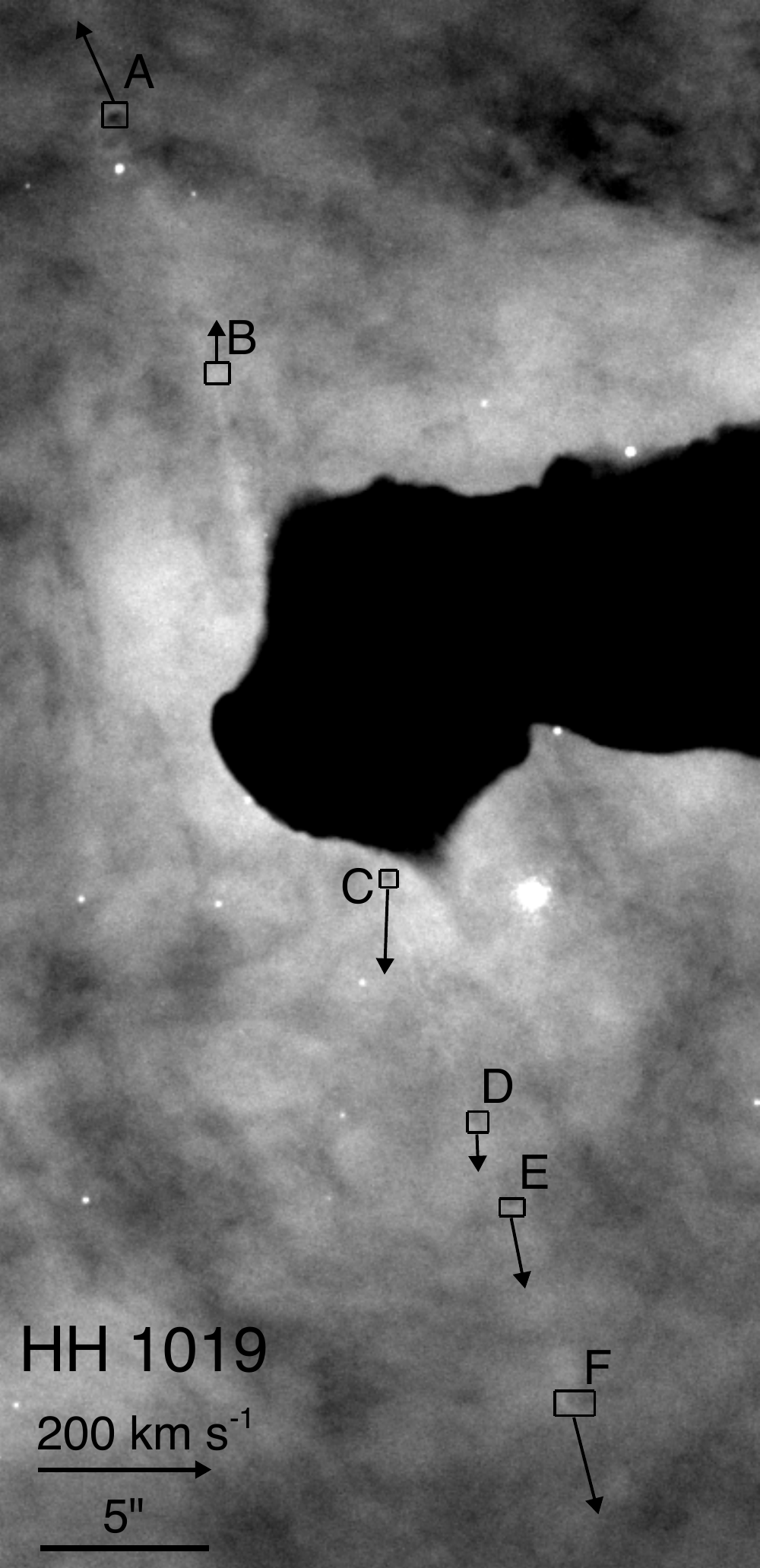}
  \caption{
    H$\alpha$ image of HH~1019 with boxes marking the jet knots used to measure proper motions.
  Vectors indicate the magnitude and direction of the motion of the silhouette knots. 
 }\label{fig:hh1019_pm} 
\end{figure}

\begin{table*}
\caption{HH~1019 knot proper motions}
\footnotesize
\begin{tabular}{lrrrrrr}
\hline\hline
Knot & $\delta x$ & $\delta y$ & v$_T$        & age   & $\tau$ & log(M$_{knot}$) \\ 
name &     mas    &     mas    & [km s$^{-1}$] & [yr] &        & M$_{\odot}$ \\ 
\hline 
HH~1019~A &  38 (3) & -86 (3) & 103 (3) & 1754 (377) &  0.2 & -4.4 \\
HH~1019~B &   0 (3) & -45 (3) &  50 (3) & 2030 (753) & 0.03 & -6.4 \\
HH~1019~C &   3 (3) &  90 (3) &  98 (4) &  532 (423) &  0.2 & -4.8 \\
HH~1019~D &  -1 (3) &  40 (3) &  43 (3) & 2912 (812) &  0.1 & -5.1 \\
HH~1019~E & -15 (3) &  74 (3) &  83 (3) & 1842 (463) &  0.1 & -4.6 \\
HH~1019~F & -26 (3) & 102 (3) & 115 (3) & 1841 (328) &  0.1 & -4.1 \\
HH~1019~G &   ...   &   ...   &   ...   &  ...       &  0.1 & -4.8 \\
HH~1019~H &   ...   &   ...   &   ...   &  ...       & 0.05 & -5.4 \\
HH~1019~I &   ...   &   ...   &   ...   &  ...       &  0.1 & -4.6 \\
HH~1019~J &   ...   &   ...   &   ...   &  ...       &  0.1 & -5.2 \\
\hline 
\end{tabular}
\label{t:pm}
\end{table*}

\section{A dusty jet seen in silhouette}\label{s:results}

Figures~\ref{fig:hh1019_intro} and \ref{fig:hh1019_hst} show HH~1019, the first dusty protostellar jet seen in silhouette.  
At least seven knots appear dark against the bright background in the H~{\sc ii} region in the H$\alpha$ image.
All lie along a single axis.
In addition, the knots are small ($\lesssim 0\farcs5$), and trace a collimated jet axis that extends nearly perpendicular to the major axis of the natal pillar.
Together, the knots trace a bipolar jet length of $\sim 40$\arcsec. 
Both HH~1019 and its natal pillar appear dark, suggesting that they lie in front of the H~{\sc ii} region created by Tr14 and Tr16. 
Nevertheless, some faint H$\alpha$ emission can be seen along the side of the jet that faces Tr16. 
Near the dust pillar, faint H$\alpha$ emission delineates the stream of material emerging into the H~{\sc ii} region, with a morphology similar to HH~901 \citep[see][]{smi10,rei13}. 
Unlike other HH jets in Carina, H$\alpha$ emission associated with the jet is not significantly brighter than the emission from the H~{\sc ii} region.

\begin{figure*}
  \centering
  $\begin{array}{cccc}
    \includegraphics[angle=0,scale=0.35]{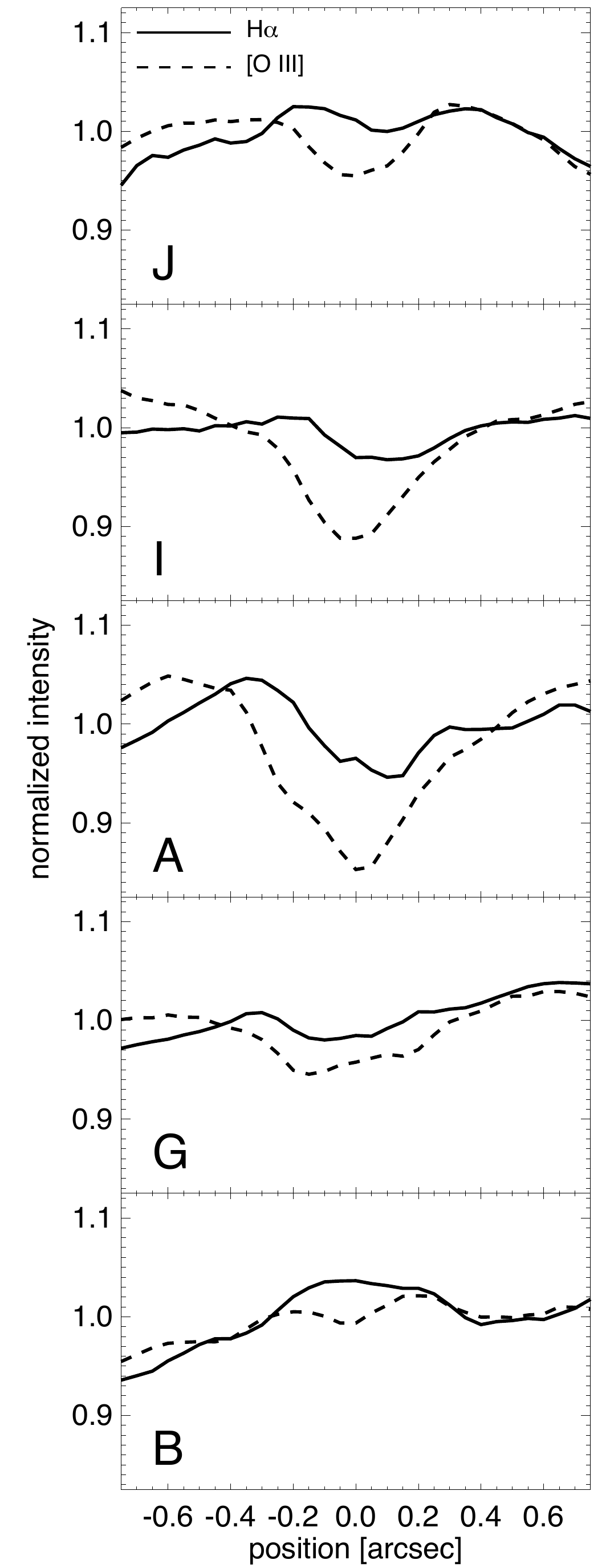} &
    \includegraphics[trim=0mm -7.5mm 0mm 0mm,angle=0,scale=0.75]{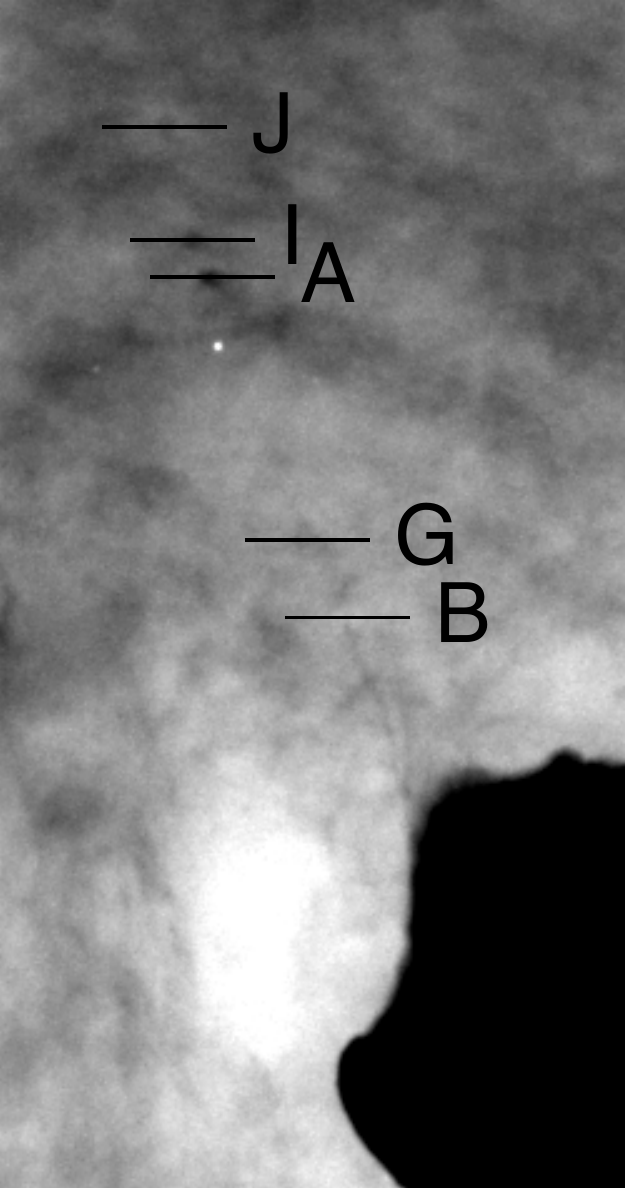} &
    \includegraphics[trim=0mm -7.5mm 0mm 0mm,angle=0,scale=0.75]{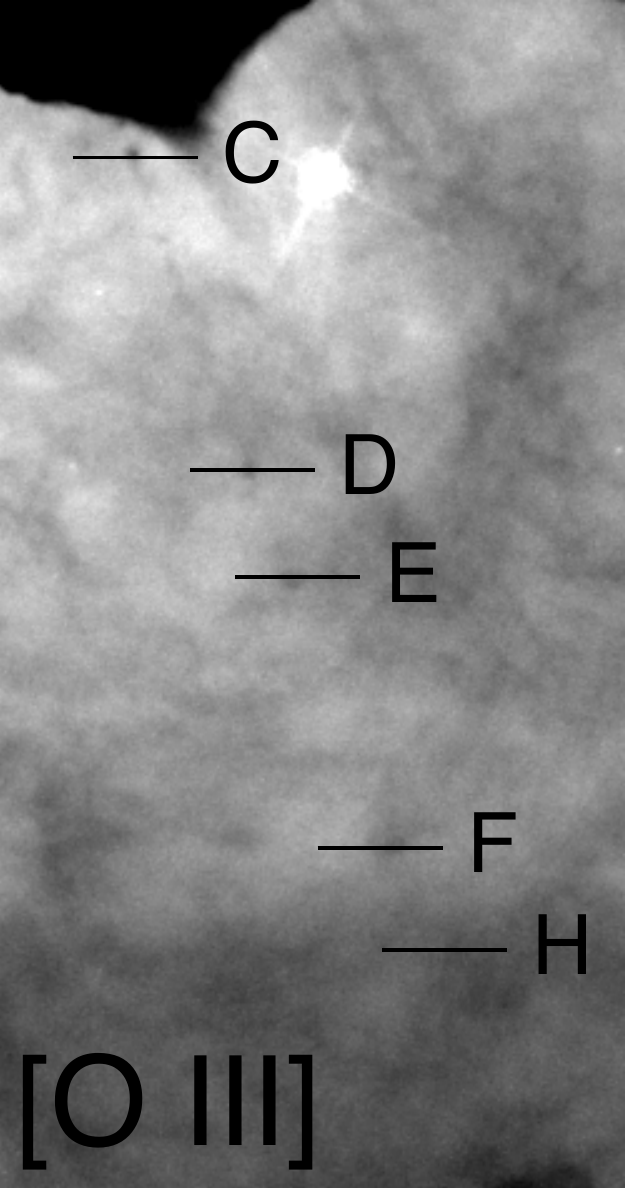} &
    \includegraphics[angle=0,scale=0.35]{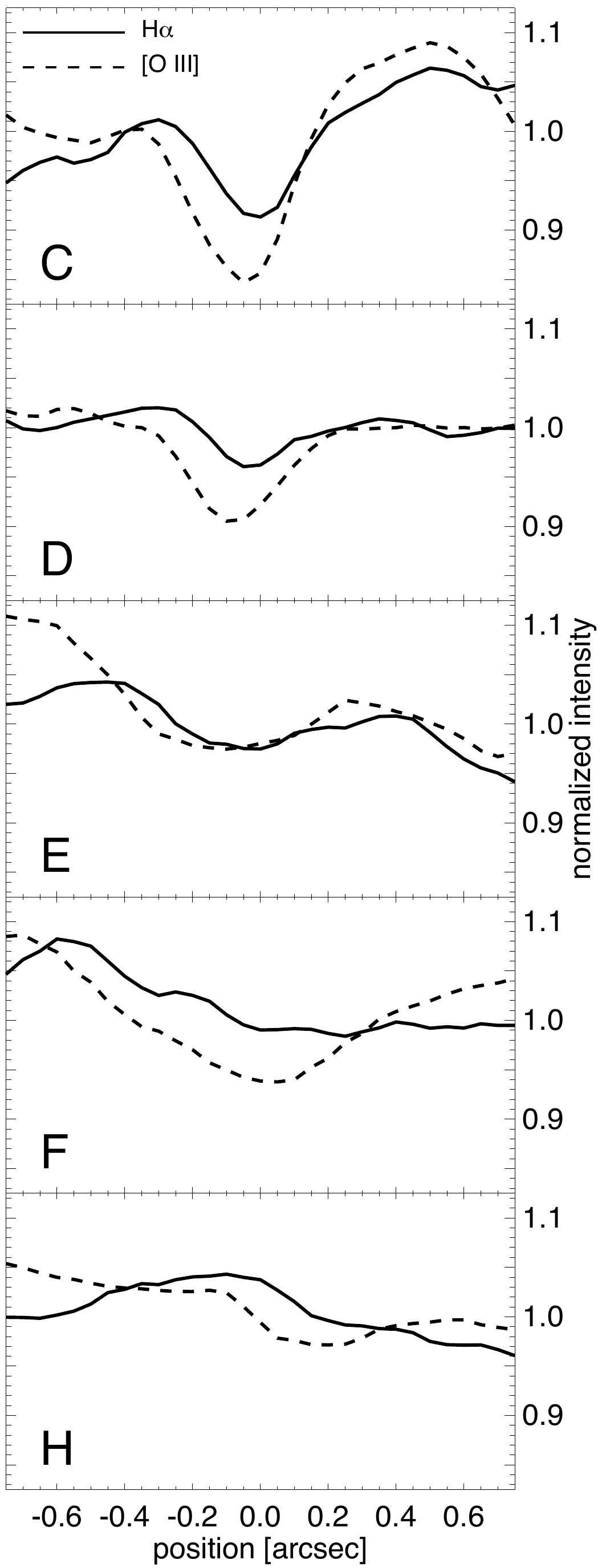} \\
 \end{array}$ 
\caption{ An \emph{HST}/ACS [O~{\sc iii}] image (center) shows dusty jets knots with lines and labels indicating the location of the intensity tracings plotted to the right and left.
  The H$\alpha$ intensity is plotted with a solid line while [O~{\sc iii}] is shown with a dashed line.
Absorption dips are deeper in [O~{\sc iii}] than H$\alpha$ due to emission from the ionization front seen at the edge of the jet (negative positions in the plots). 
 }\label{fig:hh1019_knot_tracings} 
\end{figure*}

One could imagine an extremely unlucky geometry where the relative velocity of the jet compared to the bright nebular background could produce P-Cygni-like absorption leading to the dark knots seen in the H$\alpha$ image. 
However, this cannot explain the fact that all of the knots seen in silhouette in the H$\alpha$ image are also seen in silhouette in the narrowband [O~{\sc iii}] image (see Figure~\ref{fig:hh1019_hst}).
Unlike H$\alpha$, [O~{\sc iii}] is a forbidden line, so resonant scattering cannot explain the increased absorption in jet knots. 
There is no [O~{\sc iii}] \textit{emission} associated with the jet, providing higher contrast between the knots and the background nebulosity. 
With this clearer view, we identify a clumpy stream extending off either side of the pillar head that appears to trace a nearly continuous jet.
These two streams intersect larger knots also seen in the H$\alpha$ image. 
Two additional silhouette knots in the northern limb of the jet can also be identified in the [O~{\sc iii}] image (features I and J, see Figure~\ref{fig:hh1019_knot_tracings}).

[S~{\sc ii}] emission also clearly traces the bipolar jet (see Figure~\ref{fig:hh1019_intro}).  
[S~{\sc ii}] appears smooth, but most of the small knots seen in silhouette are too small ($\lesssim 0\farcs5$) to be resolved in the seeing-limited image ($0\farcs8$).
While the coarser resolution of the seeing-limited data prevents a detailed comparison of the \emph{HST} and ground-based images, [S~{\sc ii}] emission clearly extends at least as far into the H~{\sc ii} region as the dusty knots seen in silhouette. 
Like H$\alpha$, [S~{\sc ii}] traces the jet as it emerges from the pillar head. 
Using \emph{HST} images of other externally irradiated jets in Carina, \citet{rei13} found that the [S~{\sc ii}] morphology was nearly identical to that seen in H$\alpha$, both tracing the ionized skin of the jet.
Extended [S~{\sc ii}] and H$\alpha$ may originate from the side of the jet that faces Tr16, perhaps tracing the ionized skin of the jet.

With two epochs of H$\alpha$ imaging from \emph{HST}, we can also measure the proper motions of well-defined jet knots.
A few knots clearly identified in the [O~{\sc iii}] image have inadequate contrast with the background nebulosity in the H$\alpha$ image and cannot be measured (features G,H,I,J).
We can measure proper motions of six knots and compute their transverse velocity by assuming that HH~1019 is at the same distance as $\eta$~Carinae \citep[measured to be $2300\pm50$~pc by][]{smi06b}. 
Knots in both limbs of the jet move away from the pillar with transverse velocities of $\sim 50-100$~km~s$^{-1}$ (see Figure~\ref{fig:hh1019_pm} and Table~\ref{t:pm}). 
Both kinematics and the collimated morphology in images require a driving source embedded at the head of the pillar.
However, no point source was detected in the pillar in \emph{Spitzer} surveys for young stars in Carina \citep{smi10b,pov11}. 
Several other jets propagate into the H~{\sc ii} region from opaque pillars near Tr14, but driving sources have only been detected in a few cases \citep[see, e.g.][]{rei13,rei16}.

\section{Dust in Jets}\label{s:conclusions}

Seen in silhouette against the bright background of the H~{\sc ii} region, HH~1019 is a unique protostellar outflow with the clearest, most direct evidence for dust in the jet. 
The high angular resolution of the \emph{HST} discovery images constrains the dust in HH~1019 to a narrow spine along the jet axis. 
Even without correcting for inclination, the transverse velocities we measure from the jet knots are faster than molecular outflows, more consistent with proper motions measured in protostellar jets \citep[e.g.][]{har01,bal02,yus05,kaj12,rei14}.
This strongly suggests that these knots trace the collimated jet, rather than clumpy, dusty material in a wide-angle outflow.
If this is correct, then the dust may have been launched from larger disk radii, outside the dust sublimation radius and/or entrained by the jet.

Indirect evidence for dust in jets points to an origin in a dusty disk wind. 
For example, \citet{aa11} infer the presence of dust in the DG~Tau jet from Fe depletion. 
In this case, depletion is higher in the lower velocity components that surround the fast spine of the jet \citep[see also][]{pod11}.
For a disk wind, the outflow terminal velocity is related to the Keplerian velocity at the launch point in the disk, suggesting that slower jet material originates from larger radii.
Disk rotation studies also place the origin point of these slower, outer jet layers at larger disk radii \citep[see, e.g.][]{fer06} that are more likely to reach outside the dust sublimation radius. 
Evidence for a similar, onion-like velocity structure has been observed in a few jets in Carina \citep{rei15b,rei15a}, as expected for a jet launched from a range of disk radii.
However, unlike DG~Tau, HH~1019 emerges from a deeply embedded protostar.
It is therefore possible that the jet entrained dust from the circumstellar envelope \citep[see, e.g.][]{sha08,off14}.

Bright gas-phase [Fe~{\sc ii}], as seen in many of the jets in Carina \citep{rei16}, may indicate little depletion onto grains \citep[e.g.][]{ell14}.
However, HH~1019 provides evidence that dust survives in at least one jet in Carina. 
Multi-wavelength observations of other jets in Carina suggest the coexistence of multiple excitation and ionization states, similar to the jets in Orion where \citet{smi05} detect thermal dust emission. 
The physical conditions in the jet must allow dust in HH~1019 to survive processing through shocks and external photoevaporation in the H~{\sc ii} region.

To estimate the dust mass of the silhouette knots, we measure the decrease in [O~{\sc iii}] intensity in the knot compared to the background (see Figure~\ref{fig:hh1019_knot_tracings}).
This gives a lower bound on the optical depth, $\tau$.
We do not correct for foreground emission that contributes to the flux measured across the face of the knot, so the estimated values of $\tau$ may underestimate the true optical depth. 
Following the procedure laid out in \citet{ode94} for circumstellar disks \citep[see also][]{mcc96,mcc98}, we use this $\tau$ to convert to a column density.
Assuming that the knots are spherical, with the radius measured perpendicular to the width of the jet ($\lesssim 0\farcs5$, see tracing direction in Figure~\ref{fig:hh1019_knot_tracings}) and using the dust properties assumed in \citet{ode94}, we estimate the knot masses. 
We calculate a typical knot mass of $\sim 10^{-5}$~M$_{\odot}$ \citep[including a gas-to-dust ratio of $\sim 100$ as in][see Table~\ref{t:pm}]{mcc96}.
Such extinction estimates provide a lower limit on the true dust mass, which may be orders of magnitude larger.

\citet{mou00} argue that dust destruction in shocks is far from complete. 
Higher preshock densities ($>10^4$~cm$^{-3}$) also appear to facilitate dust survival \citep{gui09}.
\citet{rei13,rei16} have argued for densities $\gtrsim 10^4$~cm$^{-3}$ for other jets in Carina based on the survival of Fe$^+$. 
This estimate is the minimum density required to shield Fe from further ionization. 
If high densities are key to dust survival in the jet, this offers the intriguing possibility that dust may survive in other jets where Fe$^+$ is shielded.

Using the estimated mass of the dusty knots, we compute the expected thermal emission from the jet.
Faint H$\alpha$ and [S~{\sc ii}] emission tracing the jet suggests some heating in the H~{\sc ii} region so that dust temperatures may be quite warm, $T \sim 100 - 1000$~K.
We calculate the expected flux as
\begin{equation}
F_{\lambda} = \frac{M_d B_{\lambda}(T) \kappa_{\lambda}} {D^2}
\end{equation}
where
$M_d$ is the estimated dust mass (derived from the [O~{\sc iii}] intensity tracings; note that the values listed in Table~\ref{t:pm} include a gas-to-dust ratio of $100$), 
$B_{\lambda}(T)$ is the Planck function, computed for a dust temperature 
$T$, 
$\kappa_{\lambda}$ is the dust opacity \citep[we assume ice-free grains,][]{oss94}, 
$D$ is the distance to the jet (2.3~kpc). 
At 10~\micron, we expect $F_{10 \micron} \gtrsim 4 \times 10^{-4}$~Jy for a dust temperature of $T=100$~K and potentially several orders of magnitude larger if the dust temperature is $T=1000$~K. 
This is smaller than the 11.7~\micron\ flux that \citet{smi05} measure in HH~513 in Orion ($F_{11.7 \micron} = 0.04$~Jy). 
Accounting for the larger distance to Carina \citep[$\gtrsim 5\times$ further than Orion, see e.g.][]{kou17} and the slightly lower opacity at 11.7~\micron\ (a factor of $\sim 2$ lower than at 10~\micron), we find that the flux of HH~1019 is a factor of $\sim 5$ lower than HH~513. 
Nevertheless, if HH~1019 is representative of other young, dusty jets, then they will be readily detectable with JWST. 

While the details of the dust-processing in HH~1019 remain uncertain, clearly there is enough dust in the jet to survive shock processing and photoevaporation in one of the more extreme parts of the Carina H~{\sc ii} region. 
Both the morphology and kinematics of HH~1019 argue that the dust is in the body of the jet. 
Spatially resolved examples like HH~1019 are therefore valuable sources to understand the origin of dust in protostellar jets.

\smallskip\smallskip\smallskip\smallskip
\noindent {\bf ACKNOWLEDGMENTS}
\smallskip
\footnotesize
MR would like to thank Ted Bergin and Keren Sharon for useful conversations. 
HH numbers are assigned by Bo Reipurth in order to
correspond with the catalogue of HH objects maintained at http://ifa.
hawaii.edu/reipurth/; see also Reipurth, B.\ \& Reiter, M.\ 2017, \textit{A General Catalog of Herbig-Haro Objects}, 3$^{rd}$. Edition, in prep.
This work is based on observations made with the NASA/ESA
{\it Hubble Space Telescope}, obtained at the Space Telescope Science
Institute, which is operated by the Association of Universities for
Research in Astronomy, Inc.\ (AURA), under NASA contract NAS5-26555. 
Support was provided by the National
Aeronautics and Space Administration (NASA) through grants AR-12155,
GO-10241, and GO-10475 from the Space Telescope Science Institute,
which is operated by AURA, Inc., under NASA contract
NAS5-26555. Additional support for this work was provided by NASA
through awards issued by JPL/Caltech as part of programs GO-3420,
GO-20452, and GO-30848.
We acknowledge the observing team for \emph{HST}/ACS [O~{\sc iii}] image: 
Mutchler, M., Livio, M., Noll, K., Levay, Z., Frattare, L.,
 Januszewski, W., Christian, C., Borders, T.

%%=============================================================================

\bibliographystyle{mnras}
\bibliography{bibliography_hh_objs}

\begin{thebibliography}{}
\makeatletter
\relax
\def\mn@urlcharsother{\let\do\@makeother \do\$\do\&\do\#\do\^\do\_\do\%\do\~}
\def\mn@doi{\begingroup\mn@urlcharsother \@ifnextchar [ {\mn@doi@}
  {\mn@doi@[]}}
\def\mn@doi@[#1]#2{\def\@tempa{#1}\ifx\@tempa\@empty \href
  {http://dx.doi.org/#2} {doi:#2}\else \href {http://dx.doi.org/#2} {#1}\fi
  \endgroup}
\def\mn@eprint#1#2{\mn@eprint@#1:#2::\@nil}
\def\mn@eprint@arXiv#1{\href {http://arxiv.org/abs/#1} {{\tt arXiv:#1}}}
\def\mn@eprint@dblp#1{\href {http://dblp.uni-trier.de/rec/bibtex/#1.xml}
  {dblp:#1}}
\def\mn@eprint@#1:#2:#3:#4\@nil{\def\@tempa {#1}\def\@tempb {#2}\def\@tempc
  {#3}\ifx \@tempc \@empty \let \@tempc \@tempb \let \@tempb \@tempa \fi \ifx
  \@tempb \@empty \def\@tempb {arXiv}\fi \@ifundefined
  {mn@eprint@\@tempb}{\@tempb:\@tempc}{\expandafter \expandafter \csname
  mn@eprint@\@tempb\endcsname \expandafter{\@tempc}}}

\bibitem[\protect\citeauthoryear{{Agra-Amboage}, {Dougados}, {Cabrit}  \&
  {Reunanen}}{{Agra-Amboage} et~al.}{2011}]{aa11}
{Agra-Amboage} V.,  {Dougados} C.,  {Cabrit} S.,   {Reunanen} J.,  2011,
  \mn@doi [\aap] {10.1051/0004-6361/201015886}, \href
  {http://adsabs.harvard.edu/abs/2011A%26A...532A..59A} {532, A59}

\bibitem[\protect\citeauthoryear{{Anderson} \& {van der Marel}}{{Anderson} \&
  {van der Marel}}{2010}]{andersonvandermarel2010}
{Anderson} J.,  {van der Marel} R.~P.,  2010, \mn@doi [\apj]
  {10.1088/0004-637X/710/2/1032}, \href
  {http://adsabs.harvard.edu/abs/2010ApJ...710.1032A} {710, 1032}

\bibitem[\protect\citeauthoryear{{Anderson} et~al.,}{{Anderson}
  et~al.}{2008a}]{anderson2008a}
{Anderson} J.,  et~al., 2008a, \mn@doi [\aj] {10.1088/0004-6256/135/6/2055},
  \href {http://adsabs.harvard.edu/abs/2008AJ....135.2055A} {135, 2055}

\bibitem[\protect\citeauthoryear{{Anderson} et~al.,}{{Anderson}
  et~al.}{2008b}]{anderson2008b}
{Anderson} J.,  et~al., 2008b, \mn@doi [\aj] {10.1088/0004-6256/135/6/2114},
  \href {http://adsabs.harvard.edu/abs/2008AJ....135.2114A} {135, 2114}

\bibitem[\protect\citeauthoryear{{Bally}, {Heathcote}, {Reipurth}, {Morse},
  {Hartigan}  \& {Schwartz}}{{Bally} et~al.}{2002}]{bal02}
{Bally} J.,  {Heathcote} S.,  {Reipurth} B.,  {Morse} J.,  {Hartigan} P.,
  {Schwartz} R.,  2002, \mn@doi [\aj] {10.1086/339837}, \href
  {http://adsabs.harvard.edu/abs/2002AJ....123.2627B} {123, 2627}

\bibitem[\protect\citeauthoryear{{Chini}, {Ward-Thompson}, {Kirk}, {Nielbock},
  {Reipurth}  \& {Sievers}}{{Chini} et~al.}{2001}]{chi01}
{Chini} R.,  {Ward-Thompson} D.,  {Kirk} J.~M.,  {Nielbock} M.,  {Reipurth} B.,
    {Sievers} A.,  2001, \mn@doi [\aap] {10.1051/0004-6361:20010097}, \href
  {http://adsabs.harvard.edu/abs/2001A%26A...369..155C} {369, 155}

\bibitem[\protect\citeauthoryear{{Currie} et~al.,}{{Currie}
  et~al.}{1996}]{cur96}
{Currie} D.~G.,  et~al., 1996, \mn@doi [\aj] {10.1086/118083}, \href
  {http://adsabs.harvard.edu/abs/1996AJ....112.1115C} {112, 1115}

\bibitem[\protect\citeauthoryear{{Ellerbroek} et~al.,}{{Ellerbroek}
  et~al.}{2014}]{ell14}
{Ellerbroek} L.~E.,  et~al., 2014, \mn@doi [\aap]
  {10.1051/0004-6361/201323092}, \href
  {http://adsabs.harvard.edu/abs/2014A%26A...563A..87E} {563, A87}

\bibitem[\protect\citeauthoryear{{Ferreira}, {Dougados}  \&
  {Cabrit}}{{Ferreira} et~al.}{2006}]{fer06}
{Ferreira} J.,  {Dougados} C.,   {Cabrit} S.,  2006, \mn@doi [\aap]
  {10.1051/0004-6361:20054231}, \href
  {http://adsabs.harvard.edu/abs/2006A%26A...453..785F} {453, 785}

\bibitem[\protect\citeauthoryear{{Gueth}, {Bachiller}  \& {Tafalla}}{{Gueth}
  et~al.}{2003}]{gue03}
{Gueth} F.,  {Bachiller} R.,   {Tafalla} M.,  2003, \mn@doi [\aap]
  {10.1051/0004-6361:20030259}, \href
  {http://adsabs.harvard.edu/abs/2003A%26A...401L...5G} {401, L5}

\bibitem[\protect\citeauthoryear{{Guillet}, {Jones}  \& {Pineau Des
  For{\^e}ts}}{{Guillet} et~al.}{2009}]{gui09}
{Guillet} V.,  {Jones} A.~P.,   {Pineau Des For{\^e}ts} G.,  2009, \mn@doi
  [\aap] {10.1051/0004-6361/200811115}, \href
  {http://adsabs.harvard.edu/abs/2009A%26A...497..145G} {497, 145}

\bibitem[\protect\citeauthoryear{{Haro}}{{Haro}}{1952}]{har52}
{Haro} G.,  1952, \mn@doi [\apj] {10.1086/145576}, \href
  {http://adsabs.harvard.edu/abs/1952ApJ...115..572H} {115, 572}

\bibitem[\protect\citeauthoryear{{Haro}}{{Haro}}{1953}]{har53}
{Haro} G.,  1953, \mn@doi [\apj] {10.1086/145669}, \href
  {http://adsabs.harvard.edu/abs/1953ApJ...117...73H} {117, 73}

\bibitem[\protect\citeauthoryear{{Hartigan}, {Morse}, {Reipurth}, {Heathcote}
  \& {Bally}}{{Hartigan} et~al.}{2001}]{har01}
{Hartigan} P.,  {Morse} J.~A.,  {Reipurth} B.,  {Heathcote} S.,   {Bally} J.,
  2001, \mn@doi [\apjl] {10.1086/323976}, \href
  {http://adsabs.harvard.edu/abs/2001ApJ...559L.157H} {559, L157}

\bibitem[\protect\citeauthoryear{{Herbig}}{{Herbig}}{1950}]{her50}
{Herbig} G.~H.,  1950, \mn@doi [\apj] {10.1086/145232}, \href
  {http://adsabs.harvard.edu/abs/1950ApJ...111...11H} {111, 11}

\bibitem[\protect\citeauthoryear{{Herbig}}{{Herbig}}{1951}]{her51}
{Herbig} G.~H.,  1951, \mn@doi [\apj] {10.1086/145440}, \href
  {http://adsabs.harvard.edu/abs/1951ApJ...113..697H} {113, 697}

\bibitem[\protect\citeauthoryear{{Kajdi{\v c}}, {Reipurth}, {Raga}, {Bally}  \&
  {Walawender}}{{Kajdi{\v c}} et~al.}{2012}]{kaj12}
{Kajdi{\v c}} P.,  {Reipurth} B.,  {Raga} A.~C.,  {Bally} J.,   {Walawender}
  J.,  2012, \mn@doi [\aj] {10.1088/0004-6256/143/5/106}, \href
  {http://adsabs.harvard.edu/abs/2012AJ....143..106K} {143, 106}

\bibitem[\protect\citeauthoryear{{Kama}, {Min}  \& {Dominik}}{{Kama}
  et~al.}{2009}]{kam09}
{Kama} M.,  {Min} M.,   {Dominik} C.,  2009, \mn@doi [\aap]
  {10.1051/0004-6361/200912068}, \href
  {http://adsabs.harvard.edu/abs/2009A%26A...506.1199K} {506, 1199}

\bibitem[\protect\citeauthoryear{{Kiminki}, {Reiter}  \& {Smith}}{{Kiminki}
  et~al.}{2016}]{kim16}
{Kiminki} M.~M.,  {Reiter} M.,   {Smith} N.,  2016, \mn@doi [\mnras]
  {10.1093/mnras/stw2019}, \href
  {http://adsabs.harvard.edu/abs/2016MNRAS.463..845K} {463, 845}

\bibitem[\protect\citeauthoryear{{Kounkel} et~al.,}{{Kounkel}
  et~al.}{2017}]{kou17}
{Kounkel} M.,  et~al., 2017, \mn@doi [\apj] {10.3847/1538-4357/834/2/142},
  \href {http://adsabs.harvard.edu/abs/2017ApJ...834..142K} {834, 142}

\bibitem[\protect\citeauthoryear{{McCaughrean} \& {O'dell}}{{McCaughrean} \&
  {O'dell}}{1996}]{mcc96}
{McCaughrean} M.~J.,  {O'dell} C.~R.,  1996, \mn@doi [\aj] {10.1086/117934},
  \href {http://adsabs.harvard.edu/abs/1996AJ....111.1977M} {111, 1977}

\bibitem[\protect\citeauthoryear{{McCaughrean} et~al.,}{{McCaughrean}
  et~al.}{1998}]{mcc98}
{McCaughrean} M.~J.,  et~al., 1998, \mn@doi [\apjl] {10.1086/311110}, \href
  {http://adsabs.harvard.edu/abs/1998ApJ...492L.157M} {492, L157}

\bibitem[\protect\citeauthoryear{{Morse}, {Kellogg}, {Bally}, {Davidson},
  {Balick}  \& {Ebbets}}{{Morse} et~al.}{2001}]{mor01}
{Morse} J.~A.,  {Kellogg} J.~R.,  {Bally} J.,  {Davidson} K.,  {Balick} B.,
  {Ebbets} D.,  2001, \mn@doi [\apjl] {10.1086/319092}, \href
  {http://adsabs.harvard.edu/abs/2001ApJ...548L.207M} {548, L207}

\bibitem[\protect\citeauthoryear{{Mouri} \& {Taniguchi}}{{Mouri} \&
  {Taniguchi}}{2000}]{mou00}
{Mouri} H.,  {Taniguchi} Y.,  2000, \mn@doi [\apjl] {10.1086/312633}, \href
  {http://adsabs.harvard.edu/abs/2000ApJ...534L..63M} {534, L63}

\bibitem[\protect\citeauthoryear{{Nisini}, {Bacciotti}, {Giannini}, {Massi},
  {Eisl{\"o}ffel}, {Podio}  \& {Ray}}{{Nisini} et~al.}{2005}]{nis05}
{Nisini} B.,  {Bacciotti} F.,  {Giannini} T.,  {Massi} F.,  {Eisl{\"o}ffel} J.,
   {Podio} L.,   {Ray} T.~P.,  2005, \mn@doi [\aap]
  {10.1051/0004-6361:20053097}, \href
  {http://adsabs.harvard.edu/abs/2005A%26A...441..159N} {441, 159}

\bibitem[\protect\citeauthoryear{{O'dell} \& {Wen}}{{O'dell} \&
  {Wen}}{1994}]{ode94}
{O'dell} C.~R.,  {Wen} Z.,  1994, \mn@doi [\apj] {10.1086/174892}, \href
  {http://adsabs.harvard.edu/abs/1994ApJ...436..194O} {436, 194}

\bibitem[\protect\citeauthoryear{{Offner} \& {Arce}}{{Offner} \&
  {Arce}}{2014}]{off14}
{Offner} S.~S.~R.,  {Arce} H.~G.,  2014, \mn@doi [\apj]
  {10.1088/0004-637X/784/1/61}, \href
  {http://adsabs.harvard.edu/abs/2014ApJ...784...61O} {784, 61}

\bibitem[\protect\citeauthoryear{{Ohlendorf}, {Preibisch}, {Gaczkowski},
  {Ratzka}, {Grellmann}  \& {McLeod}}{{Ohlendorf} et~al.}{2012}]{ohl12}
{Ohlendorf} H.,  {Preibisch} T.,  {Gaczkowski} B.,  {Ratzka} T.,  {Grellmann}
  R.,   {McLeod} A.~F.,  2012, \mn@doi [\aap] {10.1051/0004-6361/201118181},
  \href {http://adsabs.harvard.edu/abs/2012A%26A...540A..81O} {540, A81}

\bibitem[\protect\citeauthoryear{{Ossenkopf} \& {Henning}}{{Ossenkopf} \&
  {Henning}}{1994}]{oss94}
{Ossenkopf} V.,  {Henning} T.,  1994, \aap, \href
  {http://adsabs.harvard.edu/abs/1994A%26A...291..943O} {291, 943}

\bibitem[\protect\citeauthoryear{{Podio}, {Eisl{\"o}ffel}, {Melnikov}, {Hodapp}
   \& {Bacciotti}}{{Podio} et~al.}{2011}]{pod11}
{Podio} L.,  {Eisl{\"o}ffel} J.,  {Melnikov} S.,  {Hodapp} K.~W.,   {Bacciotti}
  F.,  2011, \mn@doi [\aap] {10.1051/0004-6361/201016049}, \href
  {http://adsabs.harvard.edu/abs/2011A%26A...527A..13P} {527, A13}

\bibitem[\protect\citeauthoryear{{Povich} et~al.,}{{Povich}
  et~al.}{2011}]{pov11}
{Povich} M.~S.,  et~al., 2011, \mn@doi [\apjs] {10.1088/0067-0049/194/1/14},
  \href {http://adsabs.harvard.edu/abs/2011ApJS..194...14P} {194, 14}

\bibitem[\protect\citeauthoryear{{Reiter} \& {Smith}}{{Reiter} \&
  {Smith}}{2013}]{rei13}
{Reiter} M.,  {Smith} N.,  2013, \mn@doi [\mnras] {10.1093/mnras/stt889}, \href
  {http://adsabs.harvard.edu/abs/2013MNRAS.433.2226R} {433, 2226}

\bibitem[\protect\citeauthoryear{{Reiter} \& {Smith}}{{Reiter} \&
  {Smith}}{2014}]{rei14}
{Reiter} M.,  {Smith} N.,  2014, \mn@doi [\mnras] {10.1093/mnras/stu1979},
  \href {http://adsabs.harvard.edu/abs/2014MNRAS.445.3939R} {445, 3939}

\bibitem[\protect\citeauthoryear{{Reiter}, {Smith}, {Kiminki}, {Bally}  \&
  {Anderson}}{{Reiter} et~al.}{2015a}]{rei15a}
{Reiter} M.,  {Smith} N.,  {Kiminki} M.~M.,  {Bally} J.,   {Anderson} J.,
  2015a, \mn@doi [\mnras] {10.1093/mnras/stu177}, \href
  {http://adsabs.harvard.edu/abs/2015arXiv150106564R} {448, 3429}

\bibitem[\protect\citeauthoryear{{Reiter}, {Smith}, {Kiminki}  \&
  {Bally}}{{Reiter} et~al.}{2015b}]{rei15b}
{Reiter} M.,  {Smith} N.,  {Kiminki} M.~M.,   {Bally} J.,  2015b, \mn@doi
  [\mnras] {10.1093/mnras/stv634}, \href
  {http://adsabs.harvard.edu/abs/2015MNRAS.450..564R} {450, 564}

\bibitem[\protect\citeauthoryear{{Reiter}, {Smith}  \& {Bally}}{{Reiter}
  et~al.}{2016}]{rei16}
{Reiter} M.,  {Smith} N.,   {Bally} J.,  2016, \mn@doi [\mnras]
  {10.1093/mnras/stw2296}, \href
  {http://adsabs.harvard.edu/abs/2016MNRAS.463.4344R} {463, 4344}

\bibitem[\protect\citeauthoryear{{Shadmehri} \& {Downes}}{{Shadmehri} \&
  {Downes}}{2008}]{sha08}
{Shadmehri} M.,  {Downes} T.~P.,  2008, \mn@doi [\mnras]
  {10.1111/j.1365-2966.2008.13345.x}, \href
  {http://adsabs.harvard.edu/abs/2008MNRAS.387.1318S} {387, 1318}

\bibitem[\protect\citeauthoryear{{Shirley}, {Evans}, {Rawlings}  \&
  {Gregersen}}{{Shirley} et~al.}{2000}]{shi00}
{Shirley} Y.~L.,  {Evans} II N.~J.,  {Rawlings} J.~M.~C.,   {Gregersen} E.~M.,
  2000, \mn@doi [\apjs] {10.1086/317358}, \href
  {http://adsabs.harvard.edu/abs/2000ApJS..131..249S} {131, 249}

\bibitem[\protect\citeauthoryear{{Smith}}{{Smith}}{2006}]{smi06b}
{Smith} N.,  2006, \mn@doi [\apj] {10.1086/503766}, \href
  {http://adsabs.harvard.edu/abs/2006ApJ...644.1151S} {644, 1151}

\bibitem[\protect\citeauthoryear{{Smith}, {Bally}  \& {Morse}}{{Smith}
  et~al.}{2003}]{smi03}
{Smith} N.,  {Bally} J.,   {Morse} J.~A.,  2003, \mn@doi [\apjl]
  {10.1086/375312}, \href {http://adsabs.harvard.edu/abs/2003ApJ...587L.105S}
  {587, L105}

\bibitem[\protect\citeauthoryear{{Smith}, {Stassun}  \& {Bally}}{{Smith}
  et~al.}{2005a}]{smi05b}
{Smith} N.,  {Stassun} K.~G.,   {Bally} J.,  2005a, \mn@doi [\aj]
  {10.1086/427249}, \href {http://adsabs.harvard.edu/abs/2005AJ....129..888S}
  {129, 888}

\bibitem[\protect\citeauthoryear{{Smith}, {Bally}, {Shuping}, {Morris}  \&
  {Kassis}}{{Smith} et~al.}{2005b}]{smi05}
{Smith} N.,  {Bally} J.,  {Shuping} R.~Y.,  {Morris} M.,   {Kassis} M.,  2005b,
  \mn@doi [\aj] {10.1086/432912}, \href
  {http://adsabs.harvard.edu/abs/2005AJ....130.1763S} {130, 1763}

\bibitem[\protect\citeauthoryear{{Smith}, {Bally}  \& {Walborn}}{{Smith}
  et~al.}{2010a}]{smi10}
{Smith} N.,  {Bally} J.,   {Walborn} N.~R.,  2010a, \mn@doi [\mnras]
  {10.1111/j.1365-2966.2010.16520.x}, \href
  {http://adsabs.harvard.edu/abs/2010MNRAS.405.1153S} {405, 1153}

\bibitem[\protect\citeauthoryear{{Smith} et~al.,}{{Smith}
  et~al.}{2010b}]{smi10b}
{Smith} N.,  et~al., 2010b, \mn@doi [\mnras]
  {10.1111/j.1365-2966.2010.16792.x}, \href
  {http://adsabs.harvard.edu/abs/2010MNRAS.406..952S} {406, 952}

\bibitem[\protect\citeauthoryear{{Sohn}, {Anderson}  \& {van der Marel}}{{Sohn}
  et~al.}{2012}]{sohn2012}
{Sohn} S.~T.,  {Anderson} J.,   {van der Marel} R.~P.,  2012, \mn@doi [\apj]
  {10.1088/0004-637X/753/1/7}, \href
  {http://adsabs.harvard.edu/abs/2012ApJ...753....7S} {753, 7}

\bibitem[\protect\citeauthoryear{{Yusef-Zadeh}, {Biretta}  \&
  {Wardle}}{{Yusef-Zadeh} et~al.}{2005}]{yus05}
{Yusef-Zadeh} F.,  {Biretta} J.,   {Wardle} M.,  2005, \mn@doi [\apj]
  {10.1086/428706}, \href {http://adsabs.harvard.edu/abs/2005ApJ...624..246Y}
  {624, 246}

\makeatother
\end{thebibliography}

\end{document}